# Superconductivity in silicon nanostructures

Nikolay T. Bagraev,[a*] Wolfgang Gehlhoff,[b] Leonid E. Klyachkin,[a] Anna M. Malyarenko,[a] Vladimir V. Romanov,[c] Serguey A. Rykov[c]

[a]A.F. Ioffe Physico-Technical Institute, Polytekhnicheskaya 26, St. Petersburg, 194021, Russia

[b]Institut fuer Festkoerperphysik, Technische Universitaet Berlin, Hardenbergstr.36, D-10623 Berlin, Germany

[c]Polytechnical University, Polytekhnicheskaya 29, St. Petersburg, 195251, Russia

**Abstract**

We present the findings of single-hole and pair-hole tunneling into the negative-U centres at the quantum well – δ-barrier interfaces that seems to give rise to the superconductivity observed in the silicon nanostructures prepared by short time diffusion of boron on the n-type Si(100) surface. These Si-based nanostructures represent the p-type ultra-narrow self-assembled silicon quantum wells, 2nm, confined by the δ - barriers heavily doped with boron, 3nm. The EPR and the thermo-emf studies show that the δ - barriers appear to consist of the trigonal dipole centres, $B^+$ - $B^-$, which are caused by the negative-U reconstruction of the shallow boron acceptors, $2B^0 \rightarrow B^+ + B^-$. Using the CV and thermo-emf techniques, the transport of two-dimensional holes inside SQW is demonstrated to be accompanied by single-hole tunneling through these negative-U centres that results in the superconductivity of the δ - barriers, which seems to be in frameworks of the mechanism suggested by E. Simanek and C.S. Ting (Sol.St.Commun., 32 (1979) 731; PRL, 45 (1980) 1213). The values of the correlation gaps obtained from these measurements are in a good agreement with the data derived from the temperature and magnetic field dependencies of the magnetic susceptibility, which reveal a strong diamagnetism and additionally identify the superconductor gap value.

*Keywords:* Self-assembled silicon nanostructures, Superconductivity, Magnetic susceptibility :

## 1. Introduction

Progress in silicon nanotechnology has enabled the fabrication of the high mobility silicon quantum wells (Si-QW) of the p-type confined by the δ - barriers heavily doped with boron on the n-type Si(100) surface [1]. Furthermore, the heavily boron doping appeared to assist the superconductivity in diamond [2]. Here we report that the δ - barriers seem to reveal the superconductor properties at high concentration of holes in Si-QW thereby forming the silicon nanostructures embedded in superconductor shells.

## 2. Methods

The preparation of oxide overlayers on silicon monocrystalline surfaces is known to be favourable to the generation of the excess fluxes of self-interstitials and vacancies that exhibit the predominant crystallographic orientation along a <111> and <100> axis, respectively (Fig. 1a) [3]. In the initial stage of the oxidation, thin oxide overlayer produces excess self-interstitials that are able to create small microdefects, whereas oppositely directed fluxes of vacancies give rise to their annihilation (Figs. 1a and 1b). Since the

---

[*] Corresponding author. Tel.: +7-812-2479315; fax: +7-812-2471017; e-mail: impurity.dipole@mail.ioffe.ru.

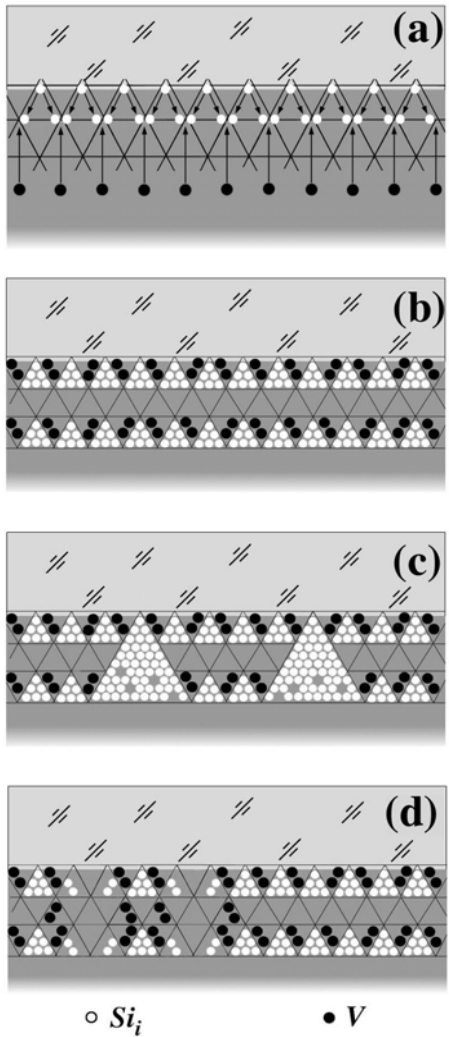

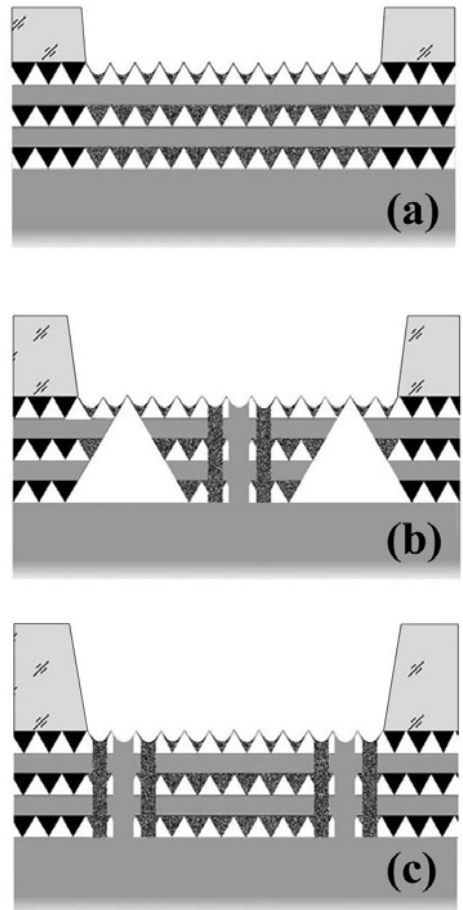

Fig. 2. A scheme of ultra-shallow diffusion profiles that consist of the longitudinal (a) and lateral (b, c) Si-QWs of the p-type prepared on the n-type Si(100) wafer in the process of the previous oxidation and subsequent short-time diffusion of boron. The atoms of boron replace the positions of vacancies thereby passivating the layers of microdefects and forming the neutral δ barriers.

Fig. 1. A scheme of self-assembled silicon quantum wells obtained by varying the thickness of the oxide overlayer prepared on the Si(100) wafer. The white and black balls label the self-interstitials and vacancies forming the excess fluxes oriented crystallographically along a <111> and <100> axis that are transformed to small microdefects (a, b). The longitudinal silicon quantum wells between the layers of microdefects are produced by performing thin oxide overlayer (b), whereas growing thick oxide overlayer results in the formation of additional lateral silicon quantum wells (d). Besides, medium and thick oxide overlayers give rise to the self-assembled microdefects of the fractal type (c).

○ $Si_i$          ● $V$

points of outgoing self-interstitials and incoming vacancies appear to be defined by the positive and negative charge states of the reconstructed silicon dangling bond [3], the dimensions of small microdefects of the self-interstitials type near the Si(100) surface have to be restricted to 2 nm. Therefore, the distribution of the microdefects created at the initial stage of the oxidation seems to represent the fractal of the Sierpinski Gasket type with the built-in longitudinal silicon quantum well (LSQW) (Fig. 1b). Then, the fractal distribution has to be reproduced by increasing the time of the oxidation process, with the $P_b$ centres as the germs for the next generation of the microdefects (Fig. 1c) [3]. The formation of thick oxide overlayer under prolonged oxidation results in however the predominant generation of vacancies by the oxidized surface, and thus, in increased decay of these microdefects, which is accompanied by the self-assembly of the lateral silicon quantum wells (LaSQW) (Fig. 1d). Although both LSQW and LaSQW embedded in the fractal system of self-assembled microdefects are of interest to be used as a basis of optically and electrically active microcavities in optoelectronics and nanoelectronics, the presence of dangling bonds at the interfaces prevents such an application. Therefore, subsequent short-time diffusion of boron would be appropriate for the passivation of dangling bonds and other defects created during previous oxidation of the Si(100) wafers thereby assisting the transformation of the arrays of microdefects in neutral δ - barriers confining the p-type high mobility ultra-narrow, 2nm, SQW that have been identified by the cyclotron resonance (CR), infrared Fourier spectroscopy and scanning tunneling microscopy (STM) methods (Figs. 2a, 2b and 2c) [1,3,4]. The EPR, SIMS thermo-emf studies have shown that the δ - barriers, 3nm, heavily doped with boron, $5 \cdot 10^{21}$ cm$^{-3}$, appear to consist of the trigonal dipole centres, B$^+$ - B$^-$ (Fig. 3), which are caused by the negative-U reconstruction of the shallow boron acceptors, $2B^0 \rightarrow B^+ + B^-$ [4]. These δ - barriers have been found to be in an excitonic insulator regime at low density of holes in SQW, with the conductance of silicon nanostructure due to the parameters of the 2D hole gas [1], whereas the density of holes more than $10^{11}$cm$^{-2}$ gives rise to the superconductivity of the δ - barriers thereby forming SQW inside the superconductor shell [5]. The present work reports on the CV, thermo-emf and magnetic susceptibility measurements that are evidence of the interplay of the electrical and magnetic properties for such superconductor LSQW.

## 3. Results

The temperature dependence of the resistivity that falls below 200 K reveals the behavior of inhomogeneous superconductor structure, with maxima and minima of peaks being in a good agreement with the thermo-emf features (Figs. 4a and 4b). The creation of each peak with decreasing temperature exhibits the logarithmic temperature dependence that appears to be caused by the Kondo-liked scattering of holes on the single fluctuations because of inhomogeneous distribution of the negative-U dipoles of boron. The positions of temperature minima of the resistivity and thermo-emf correspond to the optimal tunneling of single holes in LSQW into negative-U boron centers in the δ - barriers. This process that seems to result in high superconductor transition temperature $T_C$ observed is related to the conduction electron tunneling into the negative-U centers, which is favourable to the increase of $T_C$ in metal-silicon eutectic alloys [6, 7].

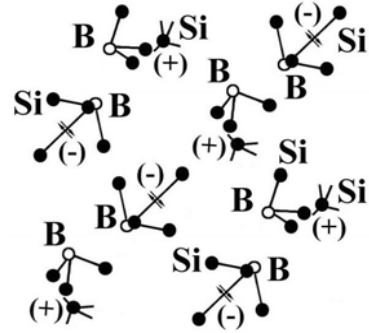

Fig. 3. Model for an elastic reconstruction of a shallow acceptor, which is accompanied by the formation of the C$_{3V}$ dipole (B$^+$ - B$^-$) centers as a result of the negative-U reaction: $2B^o \rightarrow B^+ + B^-$.

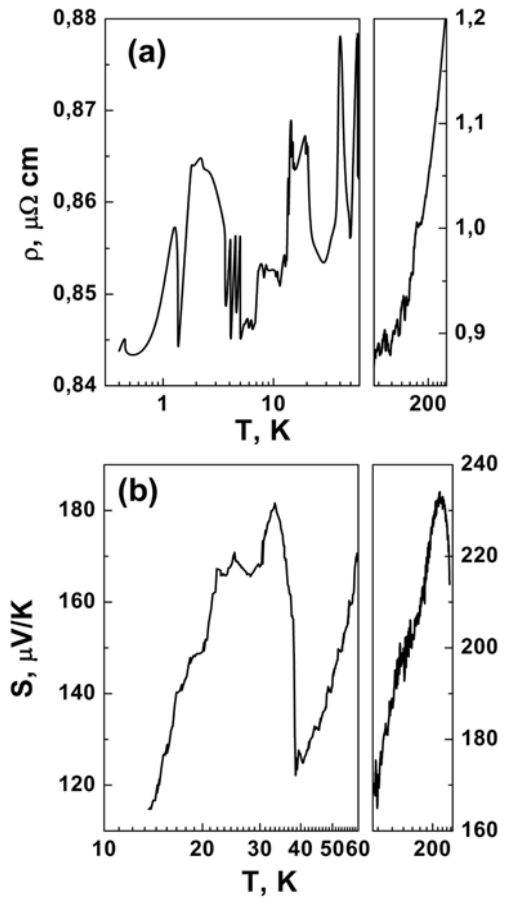

Fig. 4. Resistivity (a) and thermo-emf (Seebeck coefficient) (b) vs temperature that were observed in the ultra-shallow p+ - diffusion profile, which contains the p-type Si-QW between δ-barriers heavily doped with boron.

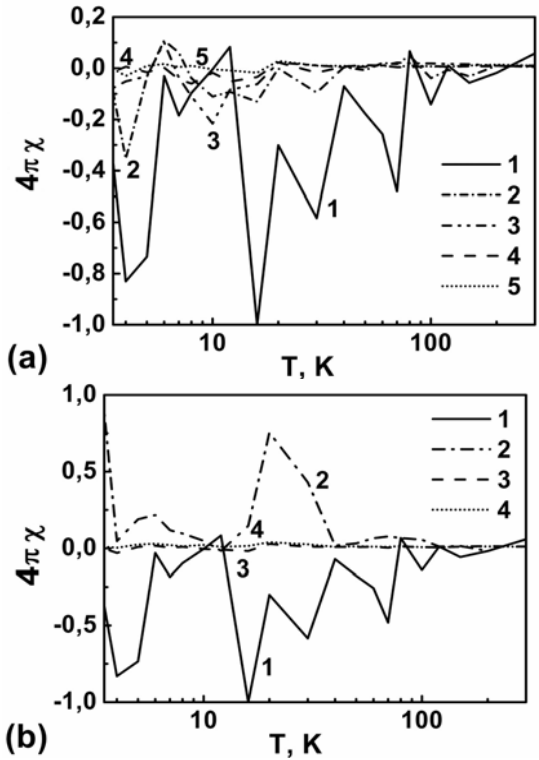

Fig. 5. Static magnetic susceptibility vs temperature that was observed in field-cooled ultra-shallow p$^+$ -diffusion profile, which contains the p-type Si-QW between δ-barriers heavily doped with boron.
(a) - diamagnetic response after field-out; 1 - 10 mT; 2 – 20 mT, 3 – 30 mT; 4 – 40 mT; 5 – 50 mT.
(b) - paramagnetic response after field-in (2, 4) and diamagnetic response after field-out (1, 3); 1,2 -10 mT, 3,4 - 50 mT.

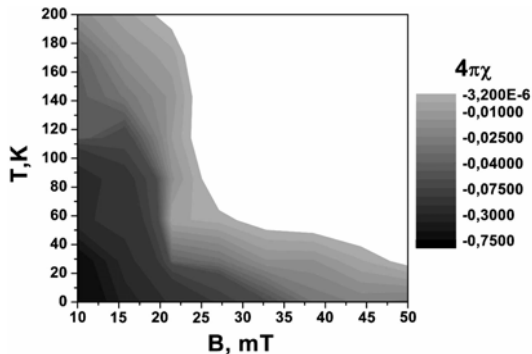

Fig. 6. Plots of static magnetic susceptibility vs temperature and magnetic field that was observed in field-cooled ultra-shallow p$^+$ -diffusion profile, which contains the p-type Si-QW between δ-barriers heavily doped with boron.

A series of correlation gaps as well as the value $\Delta$=0.022eV, $2\Delta$=3.53k$T_C$, derived from the measurements of the resistivity and thermo-emf appear to be revealed also in the temperature and magnetic field dependencies of the static magnetic susceptibility obtained by the Faraday balance method (Figs. 5a and 6). The value of temperatures corresponding to the drops of the diamagnetic response on cooling is of importance to coincide with the maxima of the resistivity and thermo-emf peaks thereby confirming the role of the charge correlations localized at the negative-U centres in the Kondo-liked scattering and the enhancement of $T_C$. Besides, the paramagnetic response observed after the field-in procedure exhibits the effect of the LSQW embedded in superconductor shell on the flux pinning processes (Fig. 5b).

Finally, the plots of the magnetic susceptibility vs temperature and magnetic field result in the value of the coherence length, $\xi$=39 nm (Fig. 6); where $\xi$= $(\Phi_0/2\pi H_{C2})^{1/2}$, $\Phi_0$ = h/2e; $H_{C2}$ is the second critical magnetic field. This value appears to be in agreement with the values of the correlation gap as well as the first critical magnetic field, $H_{C1}$=215 Oe.

## 4. Summary

Superconductivity of the p-type ultra-narrow silicon quantum wells confined by the δ - barriers heavily doped with boron has been demonstrated by the CV, thermo-emf and static magnetic susceptibility measurements. The high value of superconductor transition temperature, 150 K, seems to result from the tunneling of the single holes into the negative-U centers that represent the trigonal dipoles of boron (B$^+$ - B$^-$) forming the superconductor shell from the δ - barriers.